# Enol form of uracil in the GU "wobble" pair. NMR evidence.

D.A. Semenov, V.A. Reznikov

Novosibirsk State University

The idea of the enol form of nucleotides being responsible for the formation of some noncanonical pairs was put forward by Watson and Crick in their first paper on DNA structure [1]. In the context of this hypothesis, the GU pair may contain either the enol form of uracil and the keto form of guanine or the enol form of guanine and the keto form of uracil.

Later, in 1966, Francis Crick proposed the "wobble base pair" structure for the GU pair [2]. There are some structural data suggesting that the structure of the RNA/DNA double helix is accurately described by the earlier hypothesis. Based on X-ray structure data, Weixlbaumer et al. [3] concluded that modified uridine and guanine could pair up if uridine exists in enol form. The authors, however, interpreted this fact as an unexpected violation of the "standard GU wobble geometry". X-ray methods cannot directly identify the enol form of uracil; thus, that study [3] could only provide indirect evidence.

A direct proof of uracil existence in enol form can be obtained by using NMR spectroscopy. For the GU pair, the proton spectrum demonstrates two protons attached to heteroatoms in the 10-15 ppm region. For the "wobble base pair" structure (A), both protons taking part in the formation of hydrogen bonds between G and U must have almost equal values of chemical shifts, but this is not the case [4]. Signals from amino group protons are outside the range given above (they are in a stronger field).

For the structure based on the enol form of uracil, (C-4 atom is enolized, structure **B**), G and U could be linked by three hydrogen bonds, and the three protons involved in this should be noticeably nonequivalent, as they are positioned between different pairs of heteroatoms: oxygen-oxygen, nitrogen-nitrogen, and nitrogen-oxygen. Importantly, structure **B** is very close imitates the classic Watson-Crick base pair GC (structure **C**).

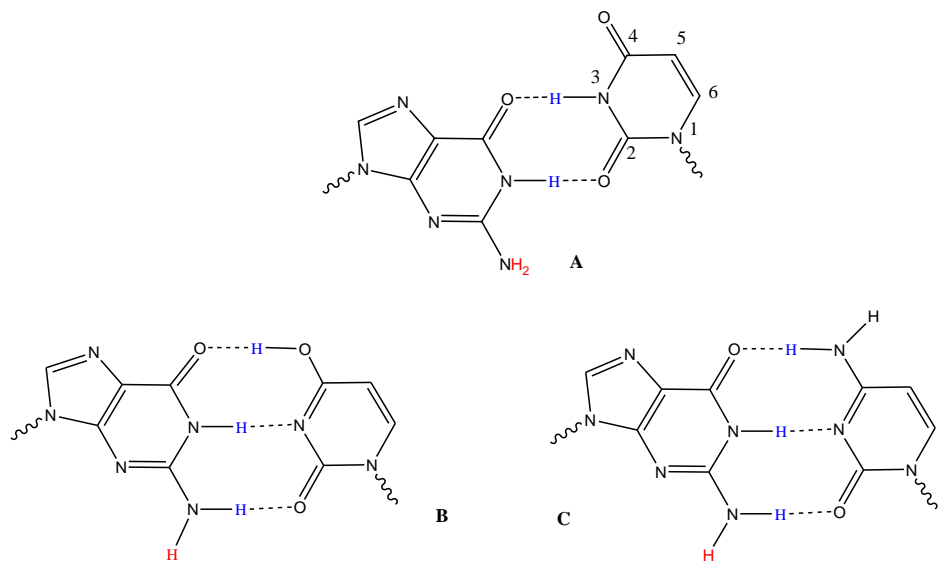

The HNN-COSY technique is used to detect hydrogen bonds between nitrogen atoms of nucleotides. We have failed to find a study in which this technique has been used exactly to investigate the GU pair, but in a number of studies the GU pair was a component of the investigated oligonucleotides. Cevec et al [5] were able to study a relatively simple oligonucleotide (with only 33 base pairs). The secondary structure of this nucleotide contains three AU pairs, eight GC pairs, and one GU pair; the loops also contain three U and one G. In the author's interpretation of the spectrum, the spectrum contains three correlated signals of the AU pairs and eight signals of the GC pairs. The GU pair is taken in the context of the wobble structure, and, hence, six more amino protons have to be observed, two of them belonging to guanine and four to uracil. The spectrum, however, shows only three signals of uracil and two of guanine. A more doubtful point of the authors' interpretation is the presence in the spectrum of an "extra" correlated pair, which is attribute to the hydrogen bond binding the terminal G1C33 pair, but this seems unlikely because of the considerable influence of the solvation of this pair (cf. [6, 7]). This suggests another interpretation: the GU pair imitates the GC pair in the spectrum due to the enolization of U, and this results in three AU, eight GC, three U and two G.

If in HNN-COSY the GU pair imitates the GC pair, results reported by Farjon at al [8] can be interpreted differently: two tautomeric forms of uracil might be observed in that study. The authors discussed in great detail the correlated signals from G and U at 12 ppm (1H) and assigned them to two non-interacting nucleotides rather than to the GU pair, i.e., they interpreted this correlation as accidental, although they correctly assigned the signal from U to the G50-U64 pair. In our opinion, the method proposed by those authors [8] enables direct observation of GU pairs, and is, thus, favorably different from the previous NMR studies. We interpret the cross peak observed in that experiment in the ≈ 190 ppm 15N region as the consequence of the presence of the enol form of uracil.

The authors would like to thank Krasova E. for her assistance in preparing this manuscript.